# On the theory of earthquakes: Paradoxical contradiction of Omori's law to the law of energy conservation

A.V. Guglielmi[1], B.I. Klain[2]

[1]*Schmidt Institute of Physics of the Earth, Russian Academy of Sciences; Bol'shaya Gruzinskaya str., 10, bld. 1, Moscow, 123242 Russia; guglielmi@mail.ru (A.G.)*
[2]*Borok Geophysical Observatory of Schmidt Institute of Physics of the Earth, Russian Academy of Sciences; klb314@mail.ru (B.K.)*

**Abstract**: After the main shock of an earthquake the aftershocks are observed. According to Omori's law, the frequency of aftershocks decreases hyperbolically over time. We noticed that, strictly speaking, Omori's law paradoxically contradicts the law of energy conservation. The contradiction is that the excitation of each aftershock consumes a finite portion of the source's energy, so that the total energy released by the source tends to infinity over time. The paradox is formally theoretical, but its analysis has proved useful. Eliminating the contradiction between Omori's law and the fundamental law of conservation of energy allowed us to further understand the nature of the phenomenological theory of aftershocks. We used the concept of deactivation of a source after the formation of a main rupture in it. We have based the theory on the original aftershock evolution equation, which has the form of a first-order linear differential equation. Two ways to eliminate the paradoxical situation are indicated.

# 1. Introduction

The review paper [1] proposed the term *earthquake triad* to denote the natural trinity of foreshocks, mainshock and aftershocks. According to the systematics of earthquakes [2], there are three types of triad, namely the *classical, mirror* and *symmetrical* triad. We will focus on classical triads, since they are characterized by an abundance of aftershocks, the evolution of which obeys the Omori law [3]. Note that aftershocks in classical triads include most tremors of tectonic origin.

130 years ago, Fusakichi Omori introduced a phenomenological parameter $k$ characterizing the state of the earthquake source, and discovered that the frequency of aftershocks $n$ decreases hyperbolically over time:

$$n(t) = \frac{k}{c+t}.\tag{1}$$

Here $k > 0$, $c > 0$, $t \geq 0$. We noticed that Omori's law (1), strictly speaking, paradoxically contradicts the law of conservation of energy. Indeed, let us introduce the source deactivation coefficient $\sigma$ after the formation of a main rupture in it, and present the law of aftershock evolution in the form of a first-order linear differential equation:

$$\dot{g} = \sigma.\tag{2}$$

Here $g = 1/n$, and the dot above the symbol means differentiation with respect to time. When $\sigma = \text{const}$ the master equation (2) is completely equivalent to Omori's law, since its solution has the form

$$n(t) = \frac{n_0}{1 + n_0 \sigma t}\tag{3}$$

and coincides with formula (1) after a simple change of notation: $\sigma = 1/k$, $n_0 = k/c$ [4]. Here $n_0 = n(0)$ is the initial condition chosen when setting up the Cauchy problem.

The total number of aftershocks excited at time $t$ is equal to

$$N(t) = n_0 \int_{t_0}^{t} \frac{dt'}{1 + n_0 \sigma t'}.\tag{4}$$

Integral (4) diverges logarithmically at $t \to +\infty$. When each aftershock is excited, a finite portion of the source energy is consumed, so that the total energy release from the source is infinite. The paradox is that equation (2), equivalent to Omori's law (1) for $\sigma = \text{const}$, describes the source as an isolated dynamic system without an influx of free energy from the outside.

We will indicate two ways to eliminate the paradoxical situation, but before that we emphasize that we fully understand the purely theoretical nature of the paradox. Real flow of aftershocks is usually interrupted by the next main shock. Apparently, this is why the paradox has not been discussed before. On the other hand, the analysis of paradoxes is always useful. We will see that analysis of the formal contradiction between Omori's law and the fundamental law of conservation of energy will allow us to better understand the essence of the phenomenological theory of aftershocks.

## 2. Eliminating the paradox
### 2.1. Omori epoch and bifurcation

It is quite obvious that the effect of Omori's law (1) is limited in time. This consideration would seem to solve the paradox, but such a solution does not satisfy us. To say that the flood of aftershocks doesn't go on forever is to say nothing. We will go another way and try to find experimentally a finite period of time, if such a period exists, during which Omori's law is exactly fulfilled in experience.

In fact, equation (2), in contrast to formula (1), allows us to take into account the non-stationary of the source, relaxing to a new state of equilibrium after the formation of the main rupture of rock continuity. To do this, we must assume that the deactivation coefficient depends on time, i.e. $\sigma = \sigma(t)$. We emphasize that Omori formula (1) excludes such an assumption. The assumption that $k = k(t)$ is unacceptable because it grossly violates spelling.

Thus, according to the evolution equation (2), the hyperbolic Omori law is not satisfied in the general case. Omori's Law can be fulfilled in individual episodes, and, very importantly, we know the strict criteria for applying Omori's Law:

$$\sigma = \text{const}. \tag{5}$$

To use criterion (5) in practice, we pose the inverse source problem [5]. It is necessary to calculate the deactivation coefficient $\sigma(t)$ based on observation data on the frequency of aftershocks $n(t)$.

The solution to the inverse problem is obvious:

$$\sigma(t) = -\frac{1}{n^2(t)} \frac{dn(t)}{dt}. \tag{6}$$

A small team of geophysicists with the participation of the authors of this paper compiled the Aftershock Atlas which provides information on the deactivation factor (6) for several dozen strong earthquakes (see, for example, the review [6]). In all cases without exception, two stages of source relaxation were observed. At the first stage, $\sigma = \text{const}$ with high accuracy. In other words, Omori's hyperbolic law is satisfied with high accuracy in the first stage. This stage has been called the *Omori epoch*. The duration of the Omori epoch varies from case to case roughly from 10 to 100 days.

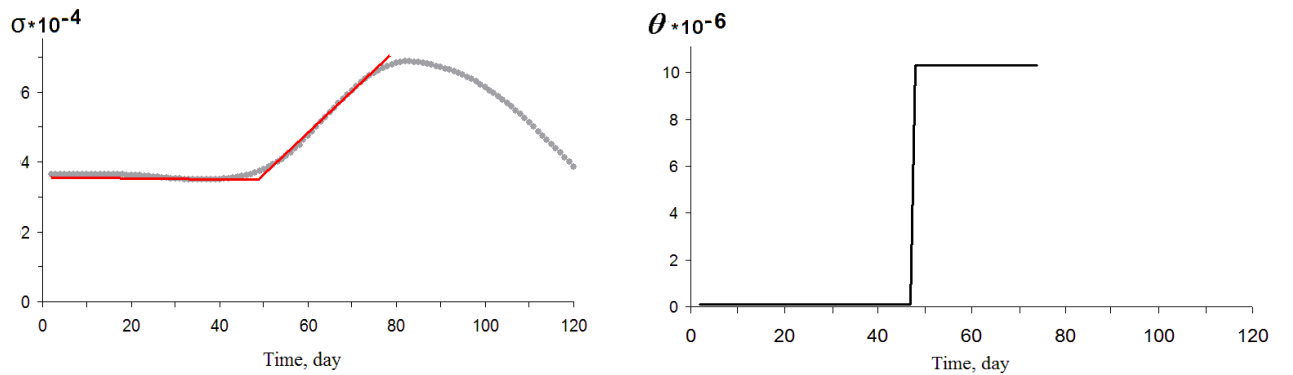

**Fig. 1.** Time dependence of deactivation coefficient (left) and its time derivative (right) after the earthquake with magnitude M = 5.9 and hypocenter depth of 6 km, which occurred on 20.07.1986 in Northern California. The duration of the Omori epoch is 48 days. The straight line segments on the left show the piecewise linear approximation of f the function $\sigma(t)$.

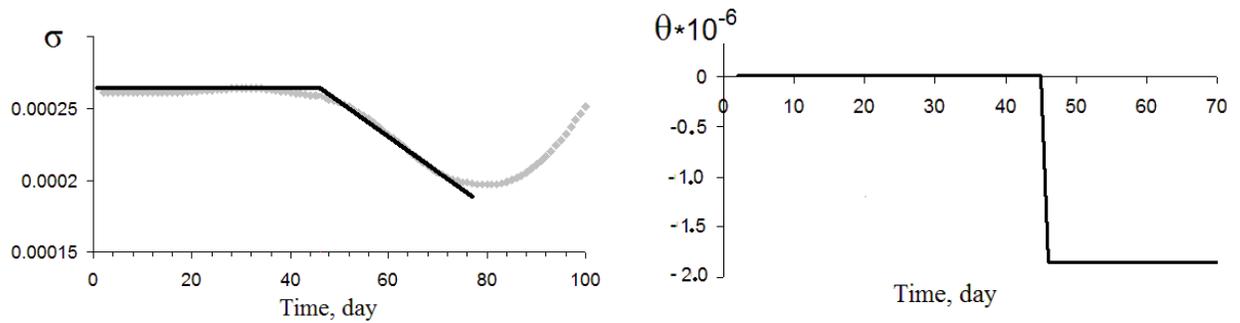

**Fig. 2.** Piecewise linear approximation of the deactivation coefficient (left) and its time derivative (right). The main shock occurred in Southern California on 16. 10.1999. The magnitude of the main shock is $M = 7.1$.

Figures 1 and 2 illustrate what has been said. On the left side of the figures we see the evolution of $\sigma(t)$ and its piecewise linear approximation. We see that the Omori epoch ($\sigma = \text{const}$) is abruptly cut off. In the second stage of the process, the deactivation factor depends on time in a rather complex way. At the end of the Omori epoch, the value of A initially increases in some cases (Figure 1), decreases in other cases (Figure 2), and behaves unpredictably thereafter. The derivative of the deactivation coefficient $\theta = d\sigma / dt$ experiences a sharp jump in the transition from the Omori epoch to the second stage of aftershock evolution (see the right parts of Figures 1 and 2). Apparently, at the moment of transition there is a bifurcation of the earthquake source [7].

The discovery of the Omori epoch is extremely important. At the end of the Omori epoch, the state of the earthquake source changes. The transition to a new state is indicated by a sharp jump of the parameter $\theta$ from zero to a finite value. This indirectly indicates that the end of the Omori epoch is accompanied by the bifurcation of the earthquake source. The transition from one mode of source deactivation to a qualitatively different mode is abrupt in the sense that its duration is much less than the duration of the Omori epoch.

Thus, a number of important conclusions can be drawn from the observations:

1. Two stages of aftershock evolution are invariably observed.

2. In the first stage, Omori's law (1) is strictly satisfied.

3. The first stage (Omori epoch) ends with a bifurcation, after which the deactivation of the source proceeds in an unpredictable way.

4. The finite duration of the first stage eliminates the paradox formulated in the Introduction.

## 2.2 Logistic equation of aftershocks

So, the discovery of the Omori epoch eliminates the paradox. Nevertheless, we will continue our analysis and give a more theoretically elegant way of dealing with the paradoxical situation. It consists in taking into account the fact that the source is not a closed passive dynamic system (2), but is an open system exchanging energy with the environment. This seems to be a fairly realistic view of the source that eliminates the paradox, but it requires instantiation in the form of an evolutionary equation more general than equation (2). In this case, the solution to the required equation must have the property $\sigma = \text{const}$ in the Omori epoch, at least, if not exactly, then in a good approximation.

At first glance, we set an exceptionally difficult problem, but its solution was found [8] thanks to the study of Faraoni [9]. Let us briefly outline the path that led us to the logistic equation of aftershocks.

It is more convenient to use here the dynamic variable $n(t)$ and to rewrite the evolution equation (2) in the form

$$\frac{dn}{dt} + \sigma n^2 = 0 \,. \tag{7}$$

In the general case $\sigma = \sigma(t)$, but at $\sigma = \text{const}$, equation (7) is exactly equivalent to the classical Omori's law (1) [4]. Faraoni [9] presented (7) in the form of Lagrange equation and drew the phase portrait of the corresponding dynamical system in phase space $(n,\dot{n})$, $n \geq 0$. The Faraoni phase portrait consists of one single phase trajectory, which has the form of half a parabola with the vertex at the point $(0,0)$. The imaging point moves upward along the parabola with deceleration to the point of stable equilibrium $(0,0)$.

This kind of "dynamical minimalism" is obviously related to the fact that the elementary master equation (7) is the simplest nonlinear evolution equation. This consideration suggested a natural generalization of the Faraoni's Lagrangian and led us to the Verhulst logistic equation

$$\frac{dn}{dt} = n(\gamma - \sigma n). \tag{8}$$

Here $\gamma$ is the second phenomenological parameter of our theory.

As a result, the phase portrait of the source has been radically enriched. The phase trajectories of the logistic equation (8) form a continuum and are located to the right and above the Faraoni trajectory. Instead of one equilibrium point, there were two, with point $(0,0)$ becoming unstable and point $(n_\infty,0)$ becoming stable. Here $n_\infty = \gamma/\sigma$.

The phase portrait analysis leads to the conclusion that the solutions to the logistic equation fall into two classes. Growing solutions correspond to segments of phase trajectories lying in the upper half of the phase space ($\dot{n} > 0$), and decreasing with time solutions in the lower half ($\dot{n} < 0$). Growing solutions are exceptionally widely used in biology, chemistry, and sociology. It seems to be the first time we have noticed that falling solutions describe aftershocks and are thus of interest in seismology.

Let us explain the geophysical meaning of the parameter $n_\infty = \gamma/\sigma$. At the end of relaxation processes, the source passes into the background seismicity mode. In this case, the energy consumption to excite tremors is completely compensated by the influx of free energy from the outside. The value of $n_\infty$ is equal to the average number of tremors in the background. It is easy to verify that the class of decreasing solutions is distinguished when posing the Cauchy problem for the logistic equation. Namely, it is enough to put $n_0 > n_\infty$.

With a sufficiently strong main shock, the strong inequality $n_0 \gg n_\infty$ is satisfied. This circumstance is of great importance. It means that in the initial stage of aftershock evolution $\gamma \ll \sigma n$ and we can neglect the first term in parentheses of equation (8). In this shortened form, the logistic equation coincides with equation (7), equivalent to the classical Omori's law (1). The ideas we have described are supported by the fact that it is at the first stage of evolution that the Omori epoch is observed, and at the same time the inequality $\gamma \ll \sigma n$ is satisfied with a sufficiently strong main shock.

### 3. Discussion

This year marks one hundred years since the death of Fusakichi Omori. He died 30 years after he formulated the first law of earthquake physics [10, 11]. Omori's Law is recognized as an outstanding achievement of seismology. The law of diminishing aftershock activity has attracted keen interest to this day.

In our opinion, the law of hyperbolic decrease of the aftershock frequency should be considered fundamental. The law is reliably fulfilled in the Omori epoch, i.e. in the first stage of evolution, which ends with the bifurcation of the source. In the second stage (after bifurcation), the evolution of aftershocks proceeds in a complex and, generally speaking, incomprehensible way. Omori's law does not work after bifurcation. If in the Omori epoch $\sigma = \text{const}$, then at the end of the epoch the deactivation coefficient becomes variable, and experience shows that the function $\sigma(t)$ is non-monotonic.

The two-stage nature of source relaxation makes it impossible to select a simple fitting formula approximating the evolution of aftershocks as a holistic process. In the first stage the fundamental Omori law $\sigma = \text{const}$ is valid, but in the second stage the deactivation factor $\sigma(t)$ behaves unpredictably, so that in general the evolution of aftershocks cannot be approximated by a simple fitting formula.

Meanwhile, formula

$$n(t) = \frac{k}{(c+t)^p} \tag{9}$$

is widely used in the literature to approximate the aftershock flux. In particular, (9) is the basis of the ETAS (*Epidemic type aftershock sequence*) theory [12, 13]. Formula (9) is sometimes called "Modified Omori function" [13], sometimes "Utsu's Law". [14]. To be fair, formula (9) belongs to Hirano, who published it in 1924 [15]. Utsu has indeed done a lot to popularize formula (9) [16-18], so it would be correct to call (9) the Hirano-Utsu formula. And we strongly disagree with Rodrigo [19], who calls formula (9) "Omori-Utsu's law". Fusakichi Omori holds the honor of discovering the fundamental law of earthquake physics (1). Omori has nothing to do with formula (9). Moreover, the fitting formula (9) cannot be considered as a law of aftershock evolution.

Indeed, Utsu and many others have convincingly shown that the frequency of aftershocks is better approximated by formula (9) than by formula (1). But that's not surprising, is it? Omori's law is one-parameter, which is clear from the differential formulation of the law (2). The Hirano-Utsu formula is a two-parameter formula. The additional fitting parameter $p$ allows for a more flexible approximation of the observed data, but nothing more.

Does the Hirano-Utsu formula (9), like the Omori formula (1), have the status of a law of earthquake physics? Our answer is definitely no. Indeed, the deactivation factor calculated by formula (9) is

$$\sigma(t) = \frac{p}{k(c+t)^{p-1}}. \tag{10}$$

At $p \neq 1$, the Omori epoch $\sigma = \text{const}$ is absent, which contradicts the observations.

Thus, formula (9) is inapplicable for describing the first stage of aftershock evolution. But it is not applicable to the second stage either, since the function (10) is monotonically increasing ($p < 1$), or monotonically decreasing ($p > 1$), while the real deactivation ratio is a non-monotonic function of time.

To conclude the discussion, let us make one remark about the logistic aftershock equation (8). The logistic equation can be arrived at in a different way from the above. It is enough to make a minimal generalization of the elementary equation (2)

$$\dot{g} + \gamma g = \sigma, \qquad (11)$$

and then make a replacement of the dynamic variable $g(t) \rightarrow n(t)$.

## 4. Summary

Strictly speaking, Omori's law in the classical formulation (1) contradicts the fundamental law of conservation of energy. We have considered two ways to eliminate the paradoxical situation. The first method is based on the results of solving the inverse problem, the essence of which is to calculate the source deactivation coefficient from the data of observing the frequency of aftershocks. Omori's law is reformulated and represented as the simplest first-order differential equation. The constancy of the deactivation coefficient is an exact criterion for the applicability of the fundamental classical Omori law. It was found that the relaxation of the source after the formation of a main rupture in it is a two-stage process. In the first stage, called the Omori epoch, the applicability criterion of the law is strictly fulfilled. The second stage begins with the bifurcation of the source, after which Omori's law is no longer applicable. The finite duration of the Omori epoch eliminates the paradox.

The second way of solving the paradox has rather theoretical significance. The elementary evolution equation, fully equivalent to Omori's law, is replaced by the logistic equation of aftershock dynamics. The logistic equation takes into account the inflow of free energy into the hearth from outside and satisfactorily describes the Omori epoch at the first stage of the source evolution.

When discussing the paradox, it was established that the well-known Hirano-Utsu formula underlying the ETAS theory is inapplicable to adequately describe the aftershock flow.

*Acknowledgements*. The Omori epoch and the bifurcation phenomenon were discovered together with A.D. Zavyalov and O.D. Zotov in the course of long-term collaboration in the experimental and theoretical study of aftershocks. To both of them we express our deepest gratitude. We thank A.L. Buchachenko, F.Z. Feygin, and A.S. Potapov for their interest in our study.

The work was carried out according to the plan of state assignments of Schmidt Institute of Physics of the Earth, Russian Academy of Sciences.